\documentclass[aps,prd,groupedaddress,showpacs,twocolumn]{revtex4}

\usepackage[latin1]{inputenc}   % only if you use accented, non-english, characters
\usepackage{latexsym}
\usepackage{graphicx}
\usepackage{hyperref}

 \providecommand{\dprod}{\! \cdot \!}%
 \providecommand{\wprod}{\! \wedge \!}
 \renewcommand{\pre}[1]{\,^#1\!}

\begin{document}

% Use the \preprint command to place your local institutional report
% number in the upper righthand corner of the title page in preprint mode.
% Multiple \preprint commands are allowed.
% Use the 'preprintnumbers' class option to override journal defaults
% to display numbers if necessary
%\preprint{}

%Title of paper
\title{Monogenic functions in 5-dimensional spacetime used as first
principle: gravitational dynamics, electromagnetism and quantum
mechanics}

% repeat the \author .. \affiliation  etc. as needed
% \email, \thanks, \homepage, \altaffiliation all apply to the current
% author. Explanatory text should go in the []'s, actual e-mail
% address or url should go in the {}'s for \email and \homepage.
% Please use the appropriate macro for each type of information

% \affiliation command applies to all authors since the last
% \affiliation command. The \affiliation command should follow the
% other information
% \affiliation can be followed by \email, \homepage, \thanks as well.
\author{José B. Almeida}
\email{bda@fisica.uminho.pt}

%\homepage[]{Your web page}
%\thanks{The author wishes to thank Frank Potter, from Sciencegems.com,
%for the enlightening discussions and corrections to the text and
%equations.}
%\altaffiliation{}
\affiliation{Universidade do Minho, Physics Department, Campus de
Gualtar, 4710-057 Braga, Portugal}

\thanks{The author wishes to thank Frank Potter, from Sciencegems.com,
for the enlightening discussions and corrections to the text and
equations.}

%Collaboration name if desired (requires use of superscriptaddress
%option in \documentclass). \noaffiliation is required (may also be
%used with the \author command).
%\collaboration can be followed by \email, \homepage, \thanks as well.
%\collaboration{}
%\noaffiliation

%\date{\today}

\begin{abstract}
Monogenic functions are functions of null vector derivative and are
here analysed in the geometric algebra of 5-dimensional spacetime,
$G_{4,1}$, in order to derive several laws of fundamental physics.
The paper introduces the working algebra and the definition of
monogenic functions, showing that these generate two 4-dimensional
spaces, one with Euclidean signature and the other one with
Minkowski signature. The equivalence conditions between the two
spaces are studied and relativistic dynamics, not entirely
coincident with Einstein's general theory of relativity, is
demonstrated. The monogenic condition is then shown to produce
Maxwell's equations and electrodynamics both classical and
quantized.
\end{abstract}

% insert suggested PACS numbers in braces on next line
\pacs{02.40.Yy; 03.65.Pm}
% insert suggested keywords - APS authors don't need to do this
%\keywords{}

%\maketitle must follow title, authors, abstract, \pacs, and \keywords
\maketitle

\section{Introduction}
Our goal is to show how the important equations of physics, such as
relativity equations and equations of quantum mechanics, can be put
under the umbrella of a common mathematical
approach\cite{Almeida05:4, Almeida04:4}. We use geometric algebra as
the framework but introduce monogenic functions with their null
derivatives in order to advance the concept.  Furthermore, we
clarify some previous work in this direction and identify the steps
to take in order to complete this ambitious project.

Since A. Einstein formulated dynamics in 4-dimensional spacetime, this
space is recognized by the vast majority of physicists as being the
best for formulating the laws of physics. However, mathematical
considerations lead to several alternative 4-D spaces. For example,
the 4-dimensional space called 4-D optics (4DO) is equivalent to the
4-D spacetime of the general theory of relativity (GTR) when the
metric is static, and therefore the geodesics of one space can be
mapped one-to-one with those of the other. Then one can choose to
work in the space that is more suitable.

In the case of a central
mass, we can examine how the Schwarzschild metric in GTR can be
transposed to 4DO. The usual form of the metric is
\begin{eqnarray}
    \mathrm{d}\tau^2 &=& \left(1-\frac{2m}{\chi} \right)\mathrm{d}t^2
    -\left(1-\frac{2m}{\chi} \right)^{-1}\mathrm{d}\chi^2 - \nonumber \\
    && - \chi^2
    \left(\mathrm{d}\theta^2 + \sin^2 \theta \mathrm{d}\varphi^2 \right);
    \label{eq:schwarz}
\end{eqnarray}
where $m$ is the spherical mass and $\chi$ is the radial coordinate,
not the distance to the centre of the mass. This form is
non-isotropic but a change of coordinates can be made that returns
the expression to isotropic form (see \citet[section
14.7]{Inverno96}):
\begin{equation}
    r=\left(\chi-m+\sqrt{\chi^2-2m \chi}\right)/2;
\end{equation}
and the new form of the metric is
\begin{eqnarray}
    \label{eq:isotropic}
    \mathrm{d}\tau^2 &=&
    \left(\frac{\displaystyle 1-\frac{m}{2r}}{\displaystyle 1+\frac{m}{2r}}\right)^2
    \mathrm{d}t^2 -
    \left(1+ \frac{m}{2r}\right)^4 \ast \\
    && \ast \left[ \mathrm{d}r^2 - r^2 \left(\mathrm{d}\theta^2
    + \sin^2 \theta \mathrm{d}\varphi^2 \right) \right]. \nonumber
\end{eqnarray}
From this equation we immediately define two coefficients, which are
called refractive index coefficients,
\begin{equation}
    \label{eq:nrn4}
    n_4 = \frac{\displaystyle 1+ \frac{m}{2r}}{\displaystyle 1-\frac{m}{2r}},
    ~~~~ n_r = \frac{\left(\displaystyle 1 + \frac{m}{2r}\right)^3}
    {\displaystyle 1-\frac{m}{2r}}.
\end{equation}

We devote the first part of this paper to deriving them from a
geometric algebra approach in a special 5D space with null
geodesics, thereby establishing that there is a 4DO Euclidean metric
space equivalent to the Schwarzschild metric space. We build upon
previous work by ourselves and by other authors about null
geodesics, regarding the condition that all material particles must
follow null geodesics of 5D space:

\begin{quotation}
{The implication of this for particles is clear: they should travel
on null 5D geodesics. This idea has recently been taken up in the
literature, and has a considerable future. It means that what we
perceive as massive particles in 4D are akin to photons in
5D.}\cite{Wesson05:1}
\end{quotation}

\begin{quotation}
{Accordingly, particles moving on null paths in 5D $(\mathrm{d}S^2 =
0)$ will appear as massive particles moving on timelike paths in 4D
$(\mathrm{d}s^2
> 0)$ \ldots}\cite{Liko03}
\end{quotation}

We actually improve on these null displacement ideas by introducing
the more fundamental monogenic condition, deriving the former from
the latter and establishing a common first principle.

\section{Some geometric algebra \label{somealg}}
Geometric algebra is not usually taught in university courses and
its presence in the literature is scarce; good reference works are
\cite{Doran03, Hestenes84, Lasenby99}. We will concentrate on the
algebra of 5-dimensional spacetime because this will be our main
working space; this algebra incorporates as subalgebras those of the
usual 3-dimensional Euclidean space, Euclidean 4-space and Minkowski
spacetime. We begin with the simpler 5D flat space and progress to a
5D spacetime of general curvature (see Appendix \ref{galgebra} for
more details.)

The geometric algebra ${G}_{4,1}$ of the hyperbolic 5-dimensional
space we consider is generated by the coordinate frame of orthonormal
basis vectors $\sigma_\alpha $ such that
\begin{eqnarray}
\label{eq:basis}
    && (\sigma_0)^2  = -1, \nonumber \\
    && (\sigma_i)^2 =1, \\
    && \sigma_\alpha \dprod \sigma_\beta  =0, \quad \alpha \neq \beta. \nonumber
    \nonumber
\end{eqnarray}
Note that the English characters i, j, k range from 1 to 4 while the
Greek characters $\alpha, \beta, \gamma$ range from 0 to 4. See the
Appendix \ref{indices} for the complete notation convention used.

Any two basis vectors can be multiplied, producing
the new entity called a bivector. This bivector
is the geometric product or, quite simply, the product, and it is
distributive. Similarly to the product of two basis vectors, the
product of three different basis vectors produces a trivector and so
forth up to the fivevector, because five is the dimension of space.

We will simplify the notation for basis vector products using
multiple indices, i.e.\ $\sigma_\alpha \sigma_\beta \equiv
\sigma_{\alpha\beta}.$ The algebra is 32-dimensional and is spanned
by the basis
\begin{itemize}
\item 1 scalar, { $1$},
\item 5 vectors, { $\sigma_\alpha$},
\item 10 bivectors (area), { $\sigma_{\alpha\beta}$},
\item 10 trivectors (volume), { $\sigma_{\alpha\beta\gamma}$},
\item 5 tetravectors (4-volume), { $\mathrm{i} \sigma_\alpha $},
\item 1 pseudoscalar (5-volume), { $\mathrm{i} \equiv
\sigma_{01234}$}.
\end{itemize}
Several elements of this basis square to unity:
\begin{equation}
\label{eq:positive}
    (\sigma_i)^2 =  (\sigma_{0i})^2=
    (\sigma_{0i j})^2 =(\mathrm{i}\sigma_0)^2 =1.
\end{equation}
The remaining basis elements square to $-1$:
\begin{equation}
    \label{eq:negative}
    (\sigma_0)^2 = (\sigma_{ij})^2 = (\sigma_{ijk})^2 =
    (\mathrm{i}\sigma_i)^2 = \mathrm{i}^2=-1.
\end{equation}
Note that the pseudoscalar $\mathrm{i}$ commutes with all the other
basis elements while being a square root of $-1$; this makes it a
very special element which can play the role of the scalar imaginary
in complex algebra.

In 5-dimensional spacetime of general curvature,
spanned by 5 coordinate frame vectors $g_\alpha$, the indices follow the
conventions set forth in Appendix \ref{indices}. We will also assume
this spacetime to be a metric space whose metric tensor is given by
\begin{equation}
\label{eq:metrictens}
    g_{\alpha \beta} = g_\alpha \dprod g_\beta;
\end{equation}
the double index is used with $g$ to denote the inner product of
frame vectors and not their geometric product. The space signature
is $(-++++)$, which amounts to saying that $g_{00} < 0$ and $g_{ii}
>0$. A reciprocal frame is defined by the condition
\begin{equation}
   \label{eq:recframe}
    g^\alpha \dprod g_\beta = {\delta^\alpha}_\beta.
\end{equation}
Defining $g^{\alpha \beta}$ as the inverse of $g_{\alpha \beta}$,
the matrix product of the two must be the identity matrix; using
Einstein's summation convention this is
\begin{equation}
    g^{\alpha \gamma} g_{\beta \gamma} = {\delta^\alpha}_\beta.
\end{equation}
Using the definition (\ref{eq:metrictens}) we have
\begin{equation}
    \left(g^{\alpha \gamma} g_\gamma \right)\dprod g_\beta =
    {\delta^\alpha}_\beta;
\end{equation}
comparing with Eq.\ (\ref{eq:recframe}) we  determine $g^\alpha$with
\begin{equation}
    g^\alpha = g^{\alpha \gamma} g_\gamma.
\end{equation}

If the coordinate frame vectors can be expressed as a linear combination of the
orthonormed ones, we have
\begin{equation}
    \label{eq:indexframemain}
    g_\alpha = {n^\beta}_\alpha \sigma_\beta,
\end{equation}
where ${n^\beta}_\alpha$ is called the \emph{refractive index
tensor} or simply the \emph{refractive index}; its 25 elements can
vary from point to point as a function of the
coordinates.\cite{Almeida04:4} When the refractive index is the
identity, we have $g_\alpha = \sigma_\alpha$ for the main or direct
frame and $g^0 = -\sigma_0$, $g^i = \sigma_i$ for the reciprocal
frame, so that Eq.\ (\ref{eq:recframe}) is verified. In this work we
will not consider spaces of general curvature but only those
satisfying condition (\ref{eq:indexframemain}).

The first use we will make of the reciprocal frame is for the
definition of two derivative operators. In flat space we define the
vector derivative
\begin{equation}
\label{eq:nabla}
    \nabla = \sigma^\alpha\partial_\alpha.
\end{equation}
It will be convenient, sometimes, to use vector derivatives in
subspaces of 5D space; these will be denoted by an upper index
before the $\nabla$ and the particular index used determines the
subspace to which the derivative applies; For instance $^m\nabla =
\sigma^m \partial_m = \sigma^1 \partial_1 + \sigma^2 \partial_2 +
\sigma^3 \partial_3.$ In 5-dimensional space it will be useful to
split the vector derivative into its time and 4-dimensional parts
\begin{equation}
    \nabla = -\sigma_0\partial_t + \sigma^i \partial_i
    = -\sigma_0\partial_t
    + \pre{i}\nabla.
\end{equation}

The second derivative operator is the covariant derivative,
sometimes called the \emph{Dirac operator}, and it is
defined in the reciprocal frame $g^\alpha$
\begin{equation}
\label{eq:covariant}
    \mathrm{D} = g^\alpha\partial_\alpha.
\end{equation}
Taking into account the definition of the reciprocal frame
(\ref{eq:recframe}), we see that the covariant derivative is also a
vector. In cases such as those we consider in this work, where there
is a refractive index, it will be possible to define both
derivatives in the same space.

We define also second order differential operators, designated
Laplacian and covariant Laplacian respectively, resulting from the
inner product of one derivative operator by itself. The square of a
vector is always a scalar and the vector derivative is no exception,
so the Laplacian is a scalar operator, which consequently acts
separately in each component of a multivector. For $4+1$ flat space
it is
\begin{equation}
    \nabla^2 = -\frac{\partial^2}{\partial t^2} + \pre{i}\nabla^2.
\end{equation}
One sees immediately that a 4-dimensional wave equation is obtained
by zeroing the Laplacian of some function
\begin{equation}
    \label{eq:4dwavemain}
    \nabla^2 \psi = \left(-\frac{\partial^2}{\partial t^2} +
    \pre{i}\nabla^2\right)\psi = 0.
\end{equation}
This procedure was used in Ref.\ \cite{Almeida05:4} for the
derivation of special relativity and will be extended here to
general curved spaces.

\section{The monogenic condition}
There is a class of functions of great importance, called
\emph{monogenic functions}\cite{Doran03}, characterized by having
null vector derivative; a function $\psi$ is monogenic in flat space
if and only if
\begin{equation}
    \label{eq:monogenic}
    \nabla \psi = 0.
\end{equation}
A monogenic function is not usually a scalar and has by necessity
null Laplacian, as can be seen by dotting Eq.\ (\ref{eq:monogenic})
with $\nabla$ on the left. We are then led to Eq.\
(\ref{eq:4dwavemain}), which can also be written as
\begin{equation}
    \label{eq:solution}
    \sum_i \partial_{ii} \psi = \partial_{00} \psi.
\end{equation}
This relation can be recognized as a wave equation in the 4-dimensional space
spanned by the $\sigma_i$ which will accept plane wave type solutions of
the general form
\begin{equation}
    \label{eq:psidef}
    \psi = \psi_0 \mathrm{e}^{\mathrm{i} (p_\alpha x^\alpha + \delta)},
\end{equation}
where $\psi_0$ is an amplitude whose characteristics we shall not
discuss for now, $\delta$ is a phase angle and $p_\alpha$ are
constants such that
\begin{equation}
    \label{eq:pnull}
    \sum_i (p_i)^2 - (p_0)^2 = 0.
\end{equation}

When working in curved spaces the monogenic condition is naturally
modified, replacing the vector derivative $\nabla$ with the
covariant derivative $\mathrm{D}$. A generalized monogenic function
is then a function that verifies the equation
\begin{equation}
\label{eq:genmonogenic}
    \mathrm{D} \psi = 0.
\end{equation}
  Similarly to what
happens in flat space, the covariant Laplacian is a scalar and a
monogenic function must verify the second order differential
equation
\begin{equation}
\label{eq:curvwave}
    \mathrm{D}^2 \psi = 0.
\end{equation}
It is possible to write a general expression for the covariant
Laplacian in terms of the metric tensor components (see
\cite[Section 2.11]{Arfken95}) but we will consider only situations
where that complete general expression is not needed.

When Eq.\ (\ref{eq:genmonogenic}) is multiplied on the left by
$\mathrm{D}$, we are applying second derivatives to the function,
but we are simultaneously applying first order derivatives to the
reciprocal frame vectors present in the definition of $\mathrm{D}$
itself. We can simplify the calculations if the variations of the
frame vectors are taken to be much slower than those of function
$\psi$ so that frame vector derivatives can be neglected. With this
approximation, the covariant Laplacian becomes $\mathrm{D}^2 =
g^{\alpha \beta}
\partial_{\alpha \beta}$ and Eq.\ (\ref{eq:curvwave}) can be written
\begin{equation}
\label{eq:curvwave2}
   g^{\alpha \beta} \partial_{\alpha \beta} \psi = 0.
\end{equation}
This equation can have a solution of the type given by Eq.\ (\ref{eq:psidef})
if again the derivatives of $p_\alpha$ are neglected. This
approximation is usually of the same order as the former one and should not be seen as a second restriction. Inserting Eq.\
(\ref{eq:psidef}) one sees that it is a solution if
\begin{equation}
    g^{\alpha \beta} p_\alpha p_\beta =0.
\end{equation}
This equation means that the square of vector $p = g^\alpha
p_\alpha$ is zero, that is, $p$ is a vector of zero length and is
called a null vector or \emph{nilpotent}. Vector $p$ is the
\emph{momentum vector} and should not be confused with 4-dimensional
conjugate momentum vectors defined below.

\section{Equivalence between 4DO and GTR spaces \label{dynamics}}

By setting the argument of $\psi$ constant in Eq.\ (\ref{eq:psidef})
and differentiating we can get the differential equation
\begin{equation}
    \label{eq:nullcond}
    p_\alpha \mathrm{d}x^\alpha =  0.
\end{equation}
The lhs can equivalently be written as the inner product of the two
vectors $p \cdot \mathrm{d}x = 0$, where $\mathrm{d}x = g_\beta
\mathrm{d}x^\beta$ is a general 5D elementary displacement. In 5D
hyperbolic space the inner product of two vectors can be null when
the vectors are perpendicular but also when the two vectors are
null. Since we have established that $p$ is a null vector, Eq.\
(\ref{eq:nullcond}) can be satisfied either by $\mathrm{d}x$ normal
to $p$ or by $(\mathrm{d}x)^2 = 0$. In the former case the condition
describes a 3-volume called wavefront and in the latter case it
describes the wave motion. Notice that the wavefronts are not
surfaces but volumes, because we are working with 4-dimensional
waves.

The condition describing 4D wave motion can be expanded as
\begin{equation}
    \label{eq:wavemotion}
    g_{\alpha \beta} \mathrm{d}x^\alpha
    \mathrm{d}x^\beta = 0.
\end{equation}
This condition effectively reduces the spatial dimension to four but
the resulting space is non-metric because all displacements have
zero length. We will remove this difficulty by considering two special
cases. First let us assume that vector $g_0$ is normal to the other
frame vectors so that all $g_{0i}$ factors are zeroed; condition
(\ref{eq:wavemotion}) becomes
\begin{equation}
\label{eq:nullspecial}
    g_{00}(\mathrm{d}x^0)^2 +
    g_{ij}\mathrm{d}x^i \mathrm{d}x^j = 0.
\end{equation}
All the terms in this equation are scalars and we are allowed to
rewrite it with $(\mathrm{d}x^0)^2$ in the lhs
\begin{equation}
\label{eq:4dometric}
    (\mathrm{d}x^0)^2 = -\frac{g_{ij}}{g_{00}}\, \mathrm{d}x^i
    \mathrm{d}x^j.
\end{equation}
We could have arrived at the same result by defining a 4-dimensional
displacement vector
\begin{equation}
\label{eq:velocity}
    \mathrm{d}x^0 v = \frac{-1}{\sqrt{g_{00}}}\, g_i \mathrm{d}x^i;
\end{equation}
and then squaring it to evaluate its length; $v$ is a unit vector
called velocity because its definition is similar to the usual
definition of 3-dimensional velocity; its components are
\begin{equation}
    v_i =  \frac{\mathrm{d}x^i}{\mathrm{d}x^0}.
\end{equation}
Being unitary, the velocity can be obtained by a rotation of the
$\sigma_4$ frame vector
\begin{equation}
    v = \tilde{R}\sigma_4 R.
\end{equation}
The rotation angle is a measure of the 3-dimensional velocity
component. A null angle corresponds to $v$ directed along $\sigma_4$
and null 3D component, while a $\pi/2$ angle corresponds to the
maximum possible 3D component. The idea that physical velocity can
be seen as the 3D component of a unitary 4D vector has been explored
in several papers but see \cite{Almeida01:4}.

Equation (\ref{eq:velocity}) projects the original 5-dimensional
space into a space with 4 dimensions, with Euclidean signature,
where an elementary displacement is given by the variation of
coordinate $x^0$. In the particular case where $g_0 = \sigma_0$ the
displacement vector simplifies to $\mathrm{d}x^0 v = g_i
\mathrm{d}x^i$ and we can see clearly that the signature is
Euclidean because the four $g_i$ have positive norm. Although it has
not been mentioned, we have assumed that none of the frame vectors
is a function of coordinate $x^0$.

Returning to Eq.\ (\ref{eq:wavemotion}) we can now impose the
condition that $g_4$ is normal to the other frame vectors in order
to isolate $(\mathrm{d}x^4)^2$ instead of $(\mathrm{d}x^0)^2$, as we
did before;
\begin{equation}
\label{eq:grmetric}
    (\mathrm{d}x^4)^2 = - \frac{g_{\mu\nu}}{g_{44}}\,
    \mathrm{d}x^\mu
    \mathrm{d}x^\nu .
\end{equation}
We have now projected onto 4-dimensional space with signature
$(+---)$, known as Minkowski signature. In order to check this
consider again the special case with $g_0 = \sigma_0$ and the
equation becomes
\begin{equation}
    (\mathrm{d}x^4)^2 = \frac{1}{g_{44}}\, (\mathrm{d}x^0)^2
    - \frac{g_{mn}}{g_{44}}\, \mathrm{d}x^m
    \mathrm{d}x^n ;
\end{equation}
the diagonal elements $g_{ii}$ are necessarily positive, which
allows a verification of Minkowski signature. Contrary to what
happened in the previous case, we cannot now obtain
$(\mathrm{d}x^4)^2$ by squaring a vector but we can do it by
consideration of the bivector
\begin{equation}
\label{eq:dx4}
    \mathrm{d}x^4 \nu  = \frac{1}{\sqrt{g_{44} g^{44}}}\, g_{\mu}g^4 \mathrm{d}x^\mu.
\end{equation}
All the products $g_{\mu}g^4$ are bivectors because we imposed $g_4$
to be normal to the other frame vectors. When $(\mathrm{d}x^4)^2$ is
evaluated by an inner product we notice that $g_0 g^4$ has positive
square while the three $g_m g^4$ have negative square, ensuring that
a Minkowski signature is obtained. Naturally we have to impose the
condition that none of the frame vectors depends on $x^4$. Bivector
$\nu$ is such that $\nu^2 = \nu \nu =1$ and it  can be obtained by a
Lorentz transformation of bivector $\sigma_{04}$.
\begin{equation}
    \nu = \tilde{T} \sigma_{04} T,
\end{equation}
where $T$ is of the form $T = \exp(B)$ and $B$ is a bivector whose
plane is normal to $\sigma_4$. Note that $T$ is a pure rotation when
the bivector plane is normal to both $\sigma_0$ and $\sigma_4$.

In special relativity it is usual to work in a space spanned by an
orthonormed frame of vectors $\gamma_\mu$ such that $(\gamma_0)^2 =
1$ and $(\gamma_m)^2 = -1$, producing the desired Minkowski
signature \cite{Doran03}. The geometric algebra of this space is
isomorphic to the even sub-algebra of $G_{4,1}$ and so the area
element $\mathrm{d}x^4 \nu$ (\ref{eq:dx4}) can be reformulated as a
vector called relativistic 4-velocity.

Equations (\ref{eq:4dometric}) and (\ref{eq:grmetric}) define two
alternative 4-dimensional spaces, those of \emph{4-dimensional
optics (4DO)}, with metric tensor $-g_{ij}/g_{00}$ and \emph{general
theory of relativity (GTR)} with metric tensor $-g_{\mu
\nu}/g_{44}$, respectively; in the former $x^0$ is an affine
parameter while in the latter it is $x^4$ that takes such role. In
fact Eq.\ (\ref{eq:grmetric}) only covers the spacelike part of GTR
space, because $(\mathrm{d}x^4)^2$ is necessarily non-negative.
Naturally there is the limitation that the frame vectors are
independent of both $x^0$ and $x^4$, equivalent to imposing a static
metric, and also that $g_{0i} = g_{\mu 4} =0$. Provided the metric
is static, the geodesics of 4DO can be mapped one-to-one with
spacelike geodesics of GTR and we can choose to work on the space
that best suits us for free fall dynamics. For a physical
interpretation of geometric relations it will frequently be
convenient to assign new designations to the 5D coordinates that
acquire the role of affine parameter in the null subspace. We will
then make the assignments $x^0 \equiv t$ and $x^4 \equiv \tau$.
Total derivatives with respect to these coordinates will also
receive a special notation: $\mathrm{d}f/\mathrm{d}t = \dot{f}$ and
$\mathrm{d}f/\mathrm{d}\tau = \check{f}$. Special units conventions
used in this paper are detailed in appendix \ref{units}.

Unless otherwise specified, we will assume that the frame vector
associated with coordinate $x^0$ is unitary and normal to all the
others, that is $g_0 = \sigma_0$ and $g_{0 i} = 0$. Recalling from
Eq.\ (\ref{eq:4dometric}), these conditions allow the definition of
4DO space with metric tensor $g_{ij}$. Although we could try a more
general approach, we would loose the possibility of interpreting
time as a line element and this, as we shall see, provides very
interesting and novel interpretations of physics equations. In many
cases it is also true that $g_4$ is normal to the other frame
vectors and we have seen that in those cases we can make metric
conversions between GTR and 4DO; it will be interesting, however, to
examine one or two situations with non-normal $g_4$ and so we leave
this possibility open.

For the moment we will concentrate on isotropic space, characterized
by orthogonal refractive index vectors $g_i$ whose norm can change
with coordinates but is the same for all vectors. Normally we relax
this condition by accepting that the three $g_m$ must have equal
norm but $g_4$ can be different. The reason for this relaxed
isotropy is found in the parallel we make with physics by assigning
dimensions $1$ to $3$ to physical space. Isotropy in a physical
sense need only be concerned with these dimensions and ignores what
happens with dimension 4. We will therefore characterize an
isotropic space by the refractive index frame $g_0 = \sigma_0$, $g_m
= n_r \sigma_m$, $g_4 = n_4 \sigma_4$. Indeed we could also accept a
non-orthogonal $g_4$ within the relaxed isotropy concept but we will
not do so for the moment.

Equation (\ref{eq:4dometric}) can now be written in terms of the
isotropic refractive indices as
\begin{equation}
    \mathrm{d}t^2 = (n_r)^2 \sum_m (\mathrm{d}x^m)^2 + (n_4 \mathrm{d}\tau)^2.
\end{equation}
Spherically symmetric static metrics play a special role; this means
that the refractive index can be expressed as functions of $r$ if we
adopt spherical coordinates. The previous equation then becomes
\begin{eqnarray}
\label{eq:spher4do}
    \mathrm{d}t^2 &=& (n_r)^2 \left[\mathrm{d}r^2 + r^2 (\mathrm{d}\theta^2
    + \sin^2 \theta \mathrm{d}\varphi^2)\right]+ \nonumber \\
    && + (n_4 \mathrm{d}\tau)^2.
\end{eqnarray}
Since we have $g_4$ normal to the other vectors we can apply metric
conversion and write the equivalent quadratic form for GTR
\begin{eqnarray}
\label{eq:sphergr}
    \mathrm{d}\tau^2 &=& \left(\frac{\mathrm{d}t}{n_4}\right)^2
    -  \left(\frac{n_r}{n_4}\right)^2 \ast \nonumber \\
    && \ast \left[\mathrm{d}r^2 + r^2 (\mathrm{d}\theta^2
    + \sin^2 \theta \mathrm{d}\varphi^2)\right].
\end{eqnarray}

As we stated in the introduction, the usual form of Schwarzschild's
metric is given by Eq.\ (\ref{eq:schwarz}) but a more interesting,
isotropic form is the one in Eq.\ (\ref{eq:isotropic}). The latter
can be compared to Eq.\ (\ref{eq:sphergr}) allowing the derivation
of the refractive indices in Eqs.\ (\ref{eq:nrn4}). These refractive
indices provide a 4DO Euclidean space equivalent to Schwarzschild
metric, allowing 4DO to be used as an alternative to GTR. Recalling
that we derived trajectories from solutions (\ref{eq:psidef}) of a
4-dimensional wave equation (\ref{eq:wavemotion}), it becomes clear
that orbits can also be seen as 4-dimensional guided waves by what
could be described as a 4-dimensional optical fibre. Modes are to be
expected in these waveguides and we shall say something about them
later on.

\section{Fermat's principle in 4 dimensions}

Fermat's principle applies to optics and states that the path
followed by a light ray is the one that makes the travel time an
extremum; usually it is the path that minimizes the time but in some
cases a ray can follow a path of maximum or stationary time. These
solutions are usually unstable, so one takes the view that light
must follow the quickest path. In Eq.\ (\ref{eq:4dometric}) we have
defined a time interval associated with a 4-dimensional elementary
displacement, which allows us to determine, by integration, a travel
time associated with displacements of any size along a given
4-dimensional path. We can then extend Fermat's principle to 4D and
impose an extremum requirement in order to select a privileged path
between any two 4D points. Taking the square root to Eq.\
(\ref{eq:4dometric})
\begin{equation}
    \mathrm{d}t = \sqrt{-\frac{g_{ij}}{g_{00}}\, \mathrm{d}x^i
    \mathrm{d}x^j}.
\end{equation}
Integrating between two points $P_1$ and $P_2$
\begin{equation}
    t = \int_{P_1}^{P_2} \sqrt{-\frac{g_{ij}}{g_{00}}\, \mathrm{d}x^i
    \mathrm{d}x^j} = \int_{P_1}^{P_2} \sqrt{-\frac{g_{ij}}{g_{00}}\, \dot{x}^i
    \dot{x}^j}\, \mathrm{d}t.
\end{equation}
In order to evaluate the previous integral one must know the
particular path linking the points by defining functions $x^i(t)$,
allowing the replacement $\mathrm{d}x^i = \dot{x}^i \mathrm{d}t$. At
this stage it is useful to define a Lagrangian
\begin{equation}
\label{eq:lagrangian}
    L =  -\frac{g_{ij}}{2 g_{00}}\, \dot{x}^i
    \dot{x}^j.
\end{equation}
The time integral can then be written
\begin{equation}
\label{eq:time}
    t = \int_{P_1}^{P_2} \sqrt{2 L}\, \mathrm{d}t.
\end{equation}

Time has to remain stationary against any small change of path;
therefore we envisage a slightly distorted path defined by functions
$x^i(t) + \varepsilon \chi^i(t)$, where $\varepsilon$ is arbitrarily
small and $ \chi^i(t)$ are functions that specify distortion. Since
the distortion must not affect the end points, the distortion
functions must vanish at those points. The time integral will now be
a function of $\varepsilon$ and we require that
\begin{equation}
    \left.
    \frac{\mathrm{d}t(\varepsilon)}{\mathrm{d}\varepsilon}\,\right|_{\varepsilon
    =0} = 0.
\end{equation}
Now, the Lagrangian (\ref{eq:lagrangian}) is a function of $x^i$,
through $g_{\alpha \beta}$ and also an explicit function of
$\dot{x}^i$. Allowing for a path change, through $\varepsilon$ makes
$t$ in Eq.\ (\ref{eq:time}) a function of $\varepsilon$
\begin{equation}
    t(\varepsilon) = \int_{P_1}^{P_2} \sqrt{2 L(x^i + \varepsilon
    \chi^i + \dot{x}^i + \varepsilon \dot{\chi}^i)}\, \mathrm{d}t.
\end{equation}
This can now be derived with respect to $\varepsilon$
\begin{eqnarray}
\label{eq:vareq}
    \left.
    \frac{\mathrm{d}t(\varepsilon)}{\mathrm{d}\varepsilon}\,\right|_{\varepsilon
    =0} &=& \left[
    \int_{P_1}^{P_2} \frac{1}{\sqrt{2L}}\right. \ast \nonumber \\
    && \ast \left. \left(\frac{\partial
    L}{\partial \dot{x}^i}\dot{\chi}^i + \frac{\partial L}{\partial
    x^i}\chi^i \right)\mathrm{d}t \right]_{\varepsilon
    =0}.
\end{eqnarray}
Note that the first term on the rhs can be written
\begin{equation}
    \int_{P_1}^{P_2} \frac{1}{\sqrt{2L}}\frac{\partial
    L}{\partial \dot{x}^i}\dot{\chi}^i \mathrm{d}t =
    \int_{P_1}^{P_2} \frac{\partial (\sqrt{2L})}{\partial \dot{x^i}}
    \dot{\chi}^i \mathrm{d}t.
\end{equation}
This can be integrated by parts
\begin{eqnarray}
    \int_{P_1}^{P_2} \frac{\partial (\sqrt{2L})}{\partial \dot{x^i}}
    \dot{\chi}^i \mathrm{d}t &=& \left[ \frac{\partial (\sqrt{2L})}
    {\partial \dot{x^i}} \chi^i\right]_{P_1}^{P_2} - \\
    && - \int_{P_1}^{P_2}
    \frac{\mathrm{d}}{ \mathrm{d} t} \left(\frac{\partial (\sqrt{2L})}
    {\partial \dot{x^i}}\right)\chi^i\mathrm{d}t. \nonumber
\end{eqnarray}
The first term on the second member is zero because $\chi^i$
vanishes for the end points; replacing in Eq.\ (\ref{eq:vareq})
\begin{eqnarray}
        \left.
    \frac{\mathrm{d}t(\varepsilon)}{\mathrm{d}\varepsilon}\,\right|_{\varepsilon
    =0} &=& \frac{1}{\sqrt{2}} \int_{P_1}^{P_2}
    \left[\frac{ \mathrm{d}}{ \mathrm{d} t} \left(- \frac{1}{\sqrt{L}}
    \frac{\partial L}{\partial \dot{x}^i}\right) \right. + \nonumber \\
    && + \left.
    \frac{1}{\sqrt{L}} \frac{\partial L}{\partial x^i} \right]
    \chi^i\mathrm{d}t.
\end{eqnarray}
The rhs must be zero for arbitrary distortion functions $\chi^i$, so
we conclude that the following set of four simultaneous equations
must be verified
\begin{equation}
\label{eq:euler}
    \frac{ \mathrm{d}}{ \mathrm{d} t} \left( \frac{1}{\sqrt{L}}
    \frac{\partial L}{\partial \dot{x}^i}\right)
    =  \frac{1}{\sqrt{L}} \frac{\partial L}{\partial x^i};
\end{equation}
these are called the Euler-Lagrange equations.

Consideration of Eqs.\ (\ref{eq:velocity}) and (\ref{eq:grmetric})
allows us to conclude that the Lagrangian defined by
(\ref{eq:lagrangian}) can also be written as $L = v^2/2$ and must
always equal $1/2$. From the Lagrangian one defines immediately the
conjugate momenta
\begin{equation}
    v_i = \frac{\partial L}{\partial \dot{x}^i} = \frac{-g_{i j}}{g_{00}} \dot{x}^j.
\end{equation}
Notice the use of the lower index ($v_i$) to represent momenta while
velocity components have an upper index ($v^i$). The conjugate
momenta are the components of the conjugate momentum vector
\begin{equation}
    v = \frac{g^i v_i}{\sqrt{-g_{00}}}
\end{equation}
and from Eq.\ (\ref{eq:recframe})
\begin{equation}
    \label{eq:momentvel}
    \sqrt{-g_{00}} v = g^i v_i = g^i g_{i j} \dot{x}^j = g_j \dot{x}^j.
\end{equation}
The conjugate momentum and velocity are the same but their
components are referred to the reciprocal and refractive index
frames, respectively.\footnote{In most cases $g_{00} = -1$, the
velocity can be conveniently written $v =g_i \dot{x}^i$ and
conjugate momenta $v_i = g_{ij} \dot{x}^j$.} Notice also that by
virtue of Eq.\ (\ref{eq:pnull}) it is also
\begin{equation}
\label{eq:4dveloc}
  v_i = \frac{p_i}{p_0}\, .
\end{equation}

The Euler-Lagrange equations (\ref{eq:euler}) can now be given a
simpler form
\begin{equation}
\label{eq:eulersimp}
    \dot{v}_i = \partial_i L.
\end{equation}
This set of four equations defines trajectories of minimum time in
4DO space as long as the frame vectors $g_\alpha$ are known
everywhere, independently of the fact that they may or may not be
referred to the orthonormed frame via a refractive index. By
definition these trajectories are the geodesics of 4DO space,
spanned by frame vectors $g_i/\sqrt{-g_{00}}$, with metric tensor
$-g_{ij}/g_{00}$.

Following an exactly similar procedure we can find trajectories
which extremize proper time, defined by taking the positive square
root of Eq.\ (\ref{eq:grmetric}). The Lagrangian is now defined by
\begin{equation}
    \mathcal{L} = -\frac{ 1}{2} \frac{g_{\mu\nu}}{g_{44}}\check{x}^\mu
    \check{x}^\nu.
\end{equation}
Consequently the conjugate momenta are
\begin{equation}
    \nu_\mu = \frac{\partial \mathcal{L}}{\partial \check{x}^\mu}
    = \frac{-g_{\mu\nu}}{g_{44}} \check{x}^\nu.
\end{equation}
From Eq.\ (\ref{eq:pnull}) we have $\nu_\mu = p_\mu/p_4$; the
associated Euler-Lagrange equations are
\begin{equation}
\label{eq:eulergr}
    \check{\nu}_\mu = \partial_\mu \mathcal{L}.
\end{equation}
"These are, by definition, spacelike geodesics of GTR with metric
tensor $-g_{\mu\nu}/g_{44}$ and we have thus defined a method for
one-to-one geodesic mapping between 4DO and spacelike GTR. Recalling
the conditions for this mapping to be valid, all the frame vectors
must be independent of both $t$ and $\tau$ and $g_0$ and $g_4$ must
be normal to the other 3 frame vectors. In tensor terms, all the
$g_{\alpha \beta}$ must be independent from $t$ and $\tau$ and
$g_{0i} = g_{\mu 4} =0$."

\section{The sources of refractive index}

The set of 4 equations (\ref{eq:eulersimp}) defines the geodesics of
4DO space; particularly in cases where there is a refractive index,
it defines trajectories of minimum time but does not tell us
anything about what produces the refractive index in the first
place. Similarly the set of equations (\ref{eq:eulergr}) defines the
geodesics of GTR space without telling us what shapes space. In
order to analyse this question we must return to the general case of
a refractive frame $g_\alpha$ without other impositions besides the
existence of a refractive index.

Considering the momentum vector
\begin{equation}
    p = p_\alpha g^\alpha = p_\alpha {n_\beta}^\alpha \sigma^\beta,
\end{equation}
with ${n_\alpha}^\gamma {n^\beta}_\gamma = \delta_\alpha^\beta$, we
will now take its time derivative. Using Eq.\ (\ref{eq:timeder})
\begin{equation}
\label{eq:dotp}
    \dot{p} = \dot{x} \dprod (\mathrm{D}p) = \dot{x} \dprod G.
\end{equation}
By a suitable choice of coordinates we can always have $g^0 =
\sigma^0$. We can then invoke the fact that for an elementary
particle in flat space the momentum vector components can be
associated with the concepts of energy, 3D momentum and rest mass as
$p = E \sigma^0 + \mathbf{p} + m \sigma^4$ (see \cite{Almeida05:1,
Almeida05:4} and Sec.\ \ref{QM}.) If this consequence is extended to
curved space and to mass distributions, we write $p = E \sigma^0 +
\mathbf{p} + m g^4$, where now $E$ is energy density, $\mathbf{p} =
p_m g^m$ is 3D momentum density and $m$ is mass density. The
previous equation then becomes
\begin{equation}
\label{eq:dynamicst}
    \dot{E} \sigma^0 +\dot{\mathbf{p}} + m \dot{g}^4 = \dot{x} \dprod G.
\end{equation}

When the Laplacian is applied to the momentum vector the result is
still necessarily a vector
\begin{equation}
    \label{eq:current}
    \mathrm{D}^2  p = S.
\end{equation}
Vector $S$ is called the \emph{sources vector} and can be expanded
into 25 terms as
\begin{equation}
\label{eq:sources}
    S = (\mathrm{D}^2 {n^\beta}_\alpha)
    \sigma_\beta p^\alpha =
    {S^\beta}_\alpha \sigma_\beta p^\alpha;
\end{equation}
where $p^\alpha = g^{\alpha \beta} p_\beta$. Tensor
${S^\alpha}_\beta$ contains the coefficients of the sources vector
and we call it the \emph{sources tensor}. The sources tensor
influences the shape of geodesics as we shall see in one
particularly important situation. One important consequence that we
don't pursue here is that by zeroing the sources vector one obtains
the wave equation $\mathrm{D}^2 p =0$, which accepts gravitational
wave solutions.

If $\sigma^0$ is normal to the other frame vectors we can write $p =
E(\sigma^0 + v)$ in the reciprocal frame, with $v$ a unit vector or
$p = E(-\sigma_0 + v)$ in the direct frame. Equation (\ref{eq:dotp})
can then be given the form
\begin{equation}
    \dot{E} (\sigma^0 + v) + E \dot{v} = {\sigma_0 + v} \dprod G.
\end{equation}
Since $G$ can have scalar and bivector components, the scalar part
must be responsible for the energy change, while the bivector part
rotates the velocity $v$. The bivector part of $G$ is generated by
$\mathrm{D} \wprod p$, which allows a simplification of the previous
equation to
\begin{equation}
    \dot{v} = v \dprod (\mathrm{D} \wprod v),
\end{equation}
if the frame vectors are independent of $t$. This equation is
exactly equivalent to the set of Euler-Lagrange equations
(\ref{eq:eulersimp}) but it was derived in a way which tells us when
to expect geodesic movement or free fall.

We will now investigate spherically symmetric solutions in isotropic
conditions defined by Eq.\ (\ref{eq:spher4do}); this means that the
refractive index can be expressed as functions of $r$. The vector
derivative in spherical coordinates is of course
\begin{eqnarray}
    \mathrm{D} &=& \frac{1}{n_r}\, \left(\sigma_r \partial_r + \frac{1}{r}\,
    \sigma_\theta \partial_\theta + \frac{1}{r \sin \theta}\, \sigma_\varphi
    \partial_\varphi \right)- \nonumber \\
    && - \sigma_t \partial_t   + \frac{1}{n_4}\, \sigma_\tau \partial_\tau.
\end{eqnarray}
The Laplacian is the inner product of $\mathrm{D}$ with itself but
the frame vectors' derivatives must be considered; all the
derivatives with respect to $r$ are zero and the others are
\begin{equation}
    \begin{array}{ll}
      \partial_\theta \sigma_r = \sigma_\theta, &
      \partial_\varphi \sigma_r = \sin \theta \sigma_\varphi, \\
      \partial_\theta \sigma_\theta = -\sigma_r, &
      \partial_\varphi \sigma_\theta = \cos \theta \sigma_\varphi, \\
      \partial_\theta \sigma_\varphi = 0, &
      \partial_\varphi \sigma_\varphi = -\sin \theta\, \sigma_r - \cos \theta\, \sigma_\theta.
    \end{array}
\end{equation}
After evaluation the curved Laplacian becomes
\begin{eqnarray}
    \label{eq:laplacradial}
    \mathrm{D}^2 &=& \frac{1}{(n_r)^2}\, \left(\partial_{rr} + \frac{2}{r}\, \partial_r -
    \frac{n'_r}{n_r}\, \partial_r  + \frac{1}{r^2}\, \partial_{\theta \theta}
     \right . + \nonumber \\
    && \left .
    +\frac{\cot \theta}{r^2}\, \partial_\theta
    + \frac{\csc^2 \theta}{r^2}\, \partial_{\varphi \varphi} \right)- \\
    && - \partial_{tt} + \frac{1}{(n_4)^2}\, \partial_{\tau \tau}.
    \nonumber
\end{eqnarray}

The search for solutions of Eq.\ (\ref{eq:current}) must necessarily
start with vanishing second member, a zero sources situation, which
one would implicitly assign to vacuum; this is a wrong assumption as
we will show. Zeroing the second member implies that the Laplacian
of both $n_r$ and $n_4$ must be zero; considering that they are
functions of $r$ we get the following equation for $n_r$
\begin{equation}
    n^{''}_r + \frac{2 n'_r}{r} - \frac{(n'_r)^2}{n_r} = 0,
\end{equation}
with general solution $n_r = b \exp(a/r)$. It is legitimate to make
$b =1$ because the refractive index must be unity at infinity. Using
this solution in Eq.\ (\ref{eq:laplacradial}) the Laplacian becomes
\begin{eqnarray}
    \mathrm{D}^2 &=& \mathrm{e}^{-a/r}\left(\partial_{rr} + \frac{2}{r}\,
    \partial_r
     + \frac{a }{r^2}\, \partial_r + \frac{1}{r^2}\,\, \partial_{\theta \theta}
     \right . +\\
    && \left .
    +\frac{\cot \theta}{r^2} \, \partial_\theta
    + \frac{\csc^2 \theta}{r^2}\, \partial_{\varphi \varphi}\right)
     - \partial_{tt} + \frac{1}{(n_4)^2}\, \partial_{\tau \tau};
     \nonumber
\end{eqnarray}
which produces the solution $n_4 = n_r $. So space must be truly
isotropic and not relaxed isotropic as we had allowed. The solution
we have found for the refractive index components in isotropic space
can correctly model Newton dynamics, which led the author to adhere
to it for some time \cite{Almeida01:4}. However if inserted into
Eq.\ (\ref{eq:grmetric}) this solution produces a GTR metric which
is verifiably in disagreement with observations; consequently it has
purely geometric significance.

The inadequacy of the isotropic solution found above for
relativistic predictions deserves some thought, so that we can
search for solutions guided by the results that are expected to have
physical significance. In the physical world we are never in a
situation of zero sources because the shape of space or the
existence of a refractive index must always be tested with a test
particle. A test particle is an abstraction corresponding to a point
mass considered so small as to have no influence on the shape of
space; in reality a point particle is a black hole in GTR, although
this fact is always overlooked. A test particle must be seen as
source of refractive index itself and its influence on the shape of
space should not be neglected in any circumstances. If this is the
case the solutions for vanishing sources vector may have only
geometric meaning, with no connection to physical reality.

The question is then what should go into the second member of Eq.\
(\ref{eq:current}) in order to find physically meaningful solutions.
If we are testing gravity we must assume some mass density to suffer
gravitational influence; this is what is usually designated as
non-interacting dust, meaning that some continuous distribution of
non-interacting particles follows the geodesics of space. Mass
density is expected to be associated with ${S^4}_4$; on the other
hand we are assuming that this mass density is very small and so we
use flat space Laplacian to evaluate it. We consequently make an
\emph{ad hoc} proposal for the sources vector in the second member
of Eq.\ (\ref{eq:current})
\begin{equation}
    \label{eq:statpart}
    S = -\nabla^2 n_4 \sigma_4.
\end{equation}
Equation (\ref{eq:current}) becomes
\begin{equation}
    \label{eq:gravitation}
    \mathrm{D}^2 \dot{x} = -\nabla^2 n_4 \sigma_4;
\end{equation}
as a result the equation for $n_r$ remains unchanged but the
equation for $n_4$ becomes
\begin{equation}
    n^{''}_4 + \frac{2 n'_4}{r} - \frac{n'_r n'_4}{n_r}
    = - n^{''}_4 + \frac{2 n'_4}{r}.
\end{equation}

When $n_r$ is given the exponential form found above, the solution
is $n_4 = \sqrt{n_r}$. This can now be entered into Eq.\
(\ref{eq:grmetric}) and the coefficients can be expanded in series
and compared to Schwarzschild's for the determination of parameter
$a$. The final solution, for a stationary mass $M$ is
\begin{equation}
    \label{eq:refind}
    n_r = \mathrm{e}^{2M/r},~~~~n_4 = \mathrm{e}^{M/r}.
\end{equation}
The equivalent GTR space is characterized by the quadratic form
\begin{equation}
    \mathrm{d}\tau^2 = \mathrm{e}^{-2M/r}\mathrm{d}t^2 - \mathrm{e}^{2M/r}\sum_m
    (\mathrm{d}x^m)^2.
\end{equation}
Expanding in series of $M/r$ the coefficients of this metric one
would find that the lower order terms are exactly the same as for
Schwarzschild's and so the predictions of the metrics are
indistinguishable for small values of the expansion variable.
\citet{Montanus01} arrives at the same solutions with a different
reasoning; the same metric is also due to Yilmaz
\cite{Yilmaz58,Yilmaz71,Ibison05}.

Equation (\ref{eq:gravitation}) can be interpreted in physical terms
as containing the essence of gravitation. When solved for
spherically symmetric solutions, as we have done, the first member
provides the definition of a stationary gravitational mass as the
factor $M$ appearing in the exponent and the second member defines
inertial mass as $\nabla^2 n_4$. Gravitational mass is defined with
recourse to some particle which undergoes gravitational influence
and is animated with velocity $v$ and inertial mass cannot be
defined without some field $n_4$ acting upon it. Complete
investigation of the sources tensor elements and their relation to
physical quantities is not yet done; it is believed that  16 terms
of this tensor have strong links with homologous elements of stress
tensor in GTR, while the others are related to electromagnetic
field.

\section{Electromagnetism in 5D spacetime \label{EM}}
Maxwell's equations can easily be written in the form of Eq.\
(\ref{eq:current}) if we don't impose the condition that $g_4$
should remain normal the other frame vectors; as we have seen in
section \ref{dynamics} this has the consequence that there will be
no GTR equivalent to the equations formulated in 4DO.

We will consider the non-orthonormed reciprocal frame defined by
\begin{equation}
    g^\mu = \sigma^\mu,~~~~ g^4 = \frac{q}{m}\, A^\mu \sigma_\mu +
    \sigma^4;
\end{equation}
where $q$ and $m$ are charge and mass densities, respectively, and
$A = A_\mu \sigma^\mu$ is the electromagnetic vector potential,
assumed to be a function of coordinates $t$ and $x^m$ but
independent of $\tau$. The associated direct frame has vectors
\begin{equation}
    g_\mu = \sigma_\mu - \frac{q}{m}\, A_\mu \sigma_4,~~~~g_4 =
    \sigma_4;
\end{equation}
and one can easily verify that Eq.\ (\ref{eq:recframe}) is obeyed.
The momentum vector in the reciprocal frame is $p = E \sigma^0 + p_m
\sigma^m + q A_\mu \sigma^\mu + m \sigma^4$ and $G$ in the second
member of Eq.\ (\ref{eq:dotp}) is $G = q \mathrm{D} A$. We will
assume $\mathrm{D} \dprod A$ to be zero, as one usually does in
electromagnetism; also $\mathrm{D}$ can be replaced by $\pre{\mu}
\nabla$ because the vector potential does not depend on $\tau$. It
is convenient to define the Faraday bivector $F =\, \pre{\mu} \nabla
A$, similarly to what is done in Ref.\ \cite{Doran03}; the dynamics
equation then becomes
\begin{equation}
    \dot{\mathbf{p}} +  q \dot{A} = q \dot{x} \dprod F;
\end{equation}
and rearranging
\begin{equation}
    \dot{\mathbf{p}} = q \dot{x} \dprod F - q \dot{A}.
\end{equation}
The first term in the second member is the Lorentz force and the
second term is due to the radiation of an accelerated charge.

Recalling the wave displacement vector Eq.\ (\ref{eq:wavedisp}) we
have now
\begin{equation}
    \mathrm{d}{x} =  \sigma_\alpha
    \mathrm{d}{x}^\alpha - \frac{q}{m}\,  A_\mu \sigma_4 \mathrm{d}{x}^\mu.
\end{equation}
This corresponds to a refractive index tensor whose non-zero terms
are
\begin{equation}
    {n^\alpha}_\alpha = 1,~~~~ {n^4}_\mu = -\frac{q}{m}\, A_\mu.
\end{equation}

According to Eq.\ (\ref{eq:sources}) the sources tensor has all
terms null except for the following
\begin{equation}
\label{eq:emsources}
    {S^4}_\mu = -\frac{q}{m}\, \mathrm{D}^2 A_\mu;
\end{equation}
where $\mathrm{D}$ is the covariant derivative given by
\begin{equation}
\label{eq:EMD}
    \mathrm{D} = g^\alpha \partial_\alpha = \sigma^\mu \partial_\mu +
    (\sigma^4 + \frac{q}{m}\, A_\mu \sigma^\mu) \partial_4.
\end{equation}
We can then define the current vector $J$ verifying
\begin{equation}
    \pre{\mu}\nabla^2 A =\, \pre{\mu} \nabla F = J,
\end{equation}
where
\begin{equation}
    J = -\frac{m}{q}\, {S^4}_\mu \sigma^\mu.
\end{equation}
Please refer to \cite[Chap.\ 7]{Doran03} or to \cite[Part
2]{Lasenby99} to see how these equations generate classical
electromagnetism, particularly how setting the current vector to
zero generates electromagnetic waves.

\section{Monogenic functions and quantum mechanics \label{QM}}
Dirac equation has been derived from the 5-dimensional monogenic
condition in previous works \cite{Almeida05:1, Almeida05:4}; the
motivation for returning to the subject here is the correction of
the electro-dynamics equation, which is incorrect in the earlier
paper \cite{Almeida05:1} and absent in the later one
\cite{Almeida05:4}. Because we are working in geometric algebra, our
quantum mechanics equations will inherit that character but the
isomorphism between the geometric algebra of 5D spacetime,
$G_{4,1}$, and complex algebra of $4 \ast 4$ matrices, $M(4,C)$,
ensures that they can be translated into the more usual Dirac matrix
formalism. The equivalence between the two formulations has been
amply demonstrated in the two references above.

Recalling the monogenic condition (\ref{eq:monogenic}), we will now
expand it into 3 terms
\begin{equation}
    (\sigma^0 \partial_0 + \sigma^m \partial_m + \sigma^4
    \partial_4) \psi =0.
\end{equation}
We have already established that this equation accepts solutions in
the form of Eq.\ (\ref{eq:psidef}) and we use that to evaluate the
derivative with respect to $x^4$
\begin{equation}
    (\sigma^0 \partial_0 + \sigma^m \partial_m - \mathrm{i} \sigma^4
    p_4) \psi =0.
\end{equation}
If the equation is multiplied by $\sigma^4$ on the left, the first 4
terms on the first member acquire bivector factors of the form
$\sigma^{4 \mu}$, the first of which, $\sigma^{40}$ squares to
unity, while the other 3 square to minus unity. Those bivectors
belong to the even sub-algebra of $G_{4,1}$, which is isomorphic the
the algebra of Minkowski spacetime, as we have already stated. It is
perfectly legitimate to replace the said bivector factors by Dirac
matrices, as was demonstrated in the above cited references. We can
then rewrite the monogenic condition as
\begin{equation}
    (\gamma^\mu \partial_\mu + \mathrm{i}p_4) \psi = 0,
\end{equation}
which can be immediately recognized as Dirac's equation if $p_4$ is
assigned to the particle's rest mass. The monogenic function given
by Eq.\ (\ref{eq:monogenic}) can then be given the usual physical
interpretation of a Dirac spinor
\begin{equation}
    \psi = \psi_0 \mathrm{e}^{\mathrm{i}(E t + \mathbf{p} \cdot
    \mathbf{x} + m \tau)};
\end{equation}
where $E$ is energy, $\mathbf{p}$ is 3-dimensional momentum and $m$
is rest mass.

In order to separate \emph{left} and \emph{right} spinor components
we use a technique adapted from Ref.\ \onlinecite{Doran03}. We
choose an arbitrary base element which squares to identity, for
instance $\sigma_4$, with which we form the two idempotents $(1 +
\sigma_4)/2$ and $(1 - \sigma_4)/2$. The name idempotents means that
they reproduce themselves when squared. These idempotents absorb any
$\sigma_4$ factor; as can be easily checked $(1 + \sigma_4) \sigma_4
= (1 + \sigma_4)$ and $(1 - \sigma_4) \sigma_4 = - (1 - \sigma_4)$.
Obviously we can decompose the wavefunction $\psi$ as
\begin{equation}
    \psi =   \psi \frac{1 + \sigma_4}{2} + \psi \frac{1 -
    \sigma_4}{2} = \psi_+ + \psi_-.
\end{equation}
This apparently trivial decomposition produces some surprising
results due to the following relations
\begin{eqnarray}
    \mathrm{e}^{\mathrm{i} \theta} (1 + \sigma_4) &=&
    (\cos \theta + \mathrm{i} \sin \theta) (1 + \sigma_4) \nonumber \\
    &=& (1 \cos \theta + \mathrm{i} \sigma_4 \sin \theta)
    (1 + \sigma_4) \\
    &=& \mathrm{e}^{\mathrm{i}\sigma_4 \theta} (1 + \sigma_4).
    \nonumber
\end{eqnarray}
and similarly
\begin{equation}
    \mathrm{e}^{\mathrm{i} \theta} (1 - \sigma_4)=
    \mathrm{e}^{-\mathrm{i} \sigma_4 \theta} (1 - \sigma_4).
\end{equation}
We could have chosen other idempotents, which would produce similar
results. The available idempotents generate an $SU(4)$ group and it
has been argued that they may be related to different elementary
particles.\cite{Almeida05:1}

Electrodynamics can now be implemented in the the same way used in
Sec.\ \ref{EM} to implement classical electromagnetism. The
monogenic condition must now be established with the covariant
derivative given by Eq.\ (\ref{eq:EMD})
\begin{equation}
    \sigma^\mu \partial_\mu \psi + \left(\sigma^4 + \frac{q}{m}\, A_\mu \sigma^\mu \right)
    \partial_4 \psi = 0.
\end{equation}
Multiplying on the left by $\sigma^4$ and taking $\partial_4 \psi =
\mathrm{i}m \psi$
\begin{equation}
    \left[\gamma^\mu (\partial_\mu + \mathrm{i} q A_\mu) + \mathrm{i}m \right]
    \psi = 0.
\end{equation}
This equation can be compared to what is found in any quantum
mechanics textbook..

It is now adequate to say a few words about quantization, which is
inherent to 5D monogenic functions. We have already seen that these
functions are 4-dimensional waves, that is, they have 3-dimensional
wavefronts normal to the direction of propagation. Whenever the
refractive index distribution traps one of these waves a
4-dimensional waveguide is produced, which has its own allowed
propagating modes. In the particular case of a central potential, be
it an atom's or a galaxy's nucleus, we expect spherical harmonic
modes, which produce the well known electron orbitals in the atom
and have unknown manifestations in a galaxy.

\section{Conclusion}
Every physicist dreams of finding a unified formulation for the
fundamental laws of physics. It is usually accepted that in order to
achieve such objective a new paradigm is needed, meaning that one
must surely step back from accepted physics principles and start
afresh from new simpler ones. Ideally one should have a small set of
principles, valid for all areas of physics, and all the important
relations should flow naturally from mathematical reasoning.

In this paper we extend a proposal previously made in that
direction, that one should accept 5-dimensional spacetime as the
adequate space to formulate the laws of physics and introduce in
this space the condition of monogeneity. We had shown in another
work that this condition is sufficient to arrive simultaneously at
special relativity and the free particle Dirac equation; here we
show that by generalizing the monogenic condition to bent spaces one
is able to obtain relativistic dynamics not entirely coincident with
GTR but also electrodynamics, both classical and quantized.

Maxwell's equations were also derived from the monogenic condition
and were unified to the equations responsible for gravitational
dynamics. The procedure is not yet entirely satisfactory, in the
sense that an \emph{ad hoc} proposal had to be made in respect to
inertial mass; in future work we hope to find a suitable formulation
for the derivation of curvature from first principles.

\begin{appendix}
\section{Indexing conventions \label{indices}}
\begin{figure}[htb]
\vspace{11pt} \centerline{\includegraphics[scale=1]{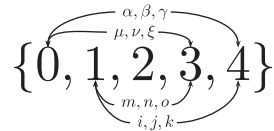}}

\caption{\label{f:indices} Indices in the range $\{0,4\}$ will be
denoted with Greek letters $\alpha, \beta, \gamma.$ Indices in the
range $\{0,3\}$ will also receive Greek letters but chosen from
$\mu, \nu, \xi.$ For indices in the range $\{1,4\}$ we will use
Latin letters $i, j, k$ and finally for indices in the range
$\{1,3\}$ we will use also Latin letters chosen from $m, n, o.$ }
\end{figure}

In this section we establish the indexing conventions used in the
paper. We deal with 5-dimensional space but we are also interested
in two of its 4-dimensional subspaces and one 3-dimensional
subspace; ideally our choice of indices should clearly identify
their ranges in order to avoid the need to specify the latter in
every equation. The diagram in Fig.\ \ref{f:indices} shows the index
naming convention used in this paper;
Einstein's summation convention will be adopted as well as the
compact notation for partial derivatives $\partial_\alpha =
\partial/\partial x^\alpha.$

\section{Non-dimensional units \label{units}}
The interpretation of $t$ and $\tau$ as time coordinates implies the
use of a scale parameter which is naturally chosen as the vacuum
speed of light $c$. We don't need to include this constant in our
equations because we can always recover time intervals, if needed,
introducing the speed of light at a later stage. We can even go a
step further and eliminate all units from our equations so that they
become pure number equations; in this way we will avoid cumbersome
constants whenever coordinates have to appear as arguments of
exponentials or trigonometric functions. We note that, at least for
the macroscopic world, physical units can all be reduced to four
fundamental ones; we can, for instance, choose length, time, mass
and electric charge as fundamental, as we could just as well have
chosen others. Measurements are then made by comparison with
standards; of course we need four standards, one for each
fundamental unit. But now note that there are four fundamental
constants: Planck constant $(\hbar)$, gravitational constant $(G)$,
speed of light in vacuum $(c)$ and proton electric charge $(e)$,
with which we can build four standards for the fundamental units.
\begin{table}[htb]
\caption{\label{t:standards}Standards for non-dimensional units'
system}
\begin{center}
\begin{tabular}{c|c|c|c}
Length & Time & Mass & Charge \\
\hline & & & \\

$\displaystyle \sqrt{\frac{G \hbar}{c^3}} $ & $\displaystyle
\sqrt{\frac{G \hbar}{c^5}} $  & $\displaystyle \sqrt{\frac{ \hbar c
}{G}} $  & $e$
\end{tabular}
\end{center}
\end{table}
Table \ref{t:standards} lists the standards of this units' system,
frequently called Planck units, which the authors prefer to
designate by non-dimensional units. In this system all the
fundamental constants, $\hbar$, $G$, $c$, $e$, become unity, a
particle's Compton frequency, defined by $\nu = mc^2/\hbar$, becomes
equal to the particle's mass and the frequent term ${GM}/({c^2 r})$
is simplified to ${M}/{r}$. We can, in fact, take all measures to be
non-dimensional, since the standards are defined with recourse to
universal constants; this will be our posture. Geometry and physics
become relations between pure numbers, vectors, bivectors, etc. and
the geometric concept of distance is needed only for graphical
representation.

\section{Some complements of geometric algebra \label{galgebra}}
In this section we expand the concepts given in Sec.\ \ref{somealg},
introducing some useful relations and definitions. Starting with the
basis elements that square to unity Eq.\ (\ref{eq:positive}),
repeated here,
\begin{equation}
    (\sigma_i)^2 =  (\sigma_{0i})^2=
    (\sigma_{0i j})^2 =(\mathrm{i}\sigma_0)^2 =1,
\end{equation}
it is easy to verify any of the above equations; suppose we want to
check that $(\sigma_{0i j})^2 = 1$. Start by expanding the square
and remove the compact notation $(\sigma_{0i j})^2 = \sigma_0
\sigma_i \sigma_j \sigma_0 \sigma_i \sigma_j$, then swap the last
$\sigma_j$ twice to bring it next to its homonymous; each swap
changes the sign, so an even number of swaps preserves the sign:
$(\sigma_{0i j})^2 = \sigma_0 \sigma_i (\sigma_j)^2 \sigma_0
\sigma_i$. From the third equation (\ref{eq:basis}) we know that the
squared vector is unity and we get successively $(\sigma_{0i j})^2 =
\sigma_0 \sigma_i \sigma_0 \sigma_i = - (\sigma_0)^2 (\sigma_i)^2 =
- (\sigma_0)^2 $; using the first equation (\ref{eq:basis}) we get
finally $(\sigma_{0i j})^2 = 1$ as desired.

The remaining basis elements square to $-1$ as can be verified in a
similar manner, Eq.\ (\ref{eq:negative}):
\begin{equation}
    (\sigma_0)^2 = (\sigma_{ij})^2 = (\sigma_{ijk})^2 =
    (\mathrm{i}\sigma_i)^2 = \mathrm{i}^2=-1.
\end{equation}
Note that the pseudoscalar $\mathrm{i}$ commutes with all the other
basis elements while being a square root of $-1$; this makes it a
very special element which can play the role of the scalar imaginary
in complex algebra.

We can now address the geometric product of any two vectors $a =
a^\alpha \sigma_\alpha$ and $b = b^\beta \sigma_\beta$ making use of
the distributive property
\begin{equation}
    ab = \left(-a^0 b^0 + \sum_i a^i b^i \right) + \sum_{\alpha \neq \beta}
    a^\alpha b^\beta \sigma_{\alpha \beta};
\end{equation}
and we notice it can be decomposed into a symmetric part, a scalar
called the inner or interior product, and an anti-symmetric part, a
bivector called the outer or exterior product.
\begin{equation}
    ab = a \dprod b + a \wprod b,~~~~ ba = a \dprod b - a \wprod b.
\end{equation}
Reversing the definition one can write inner and outer products as
\begin{equation}
    a \dprod b = \frac{1}{2}\, (ab + ba),~~~~ a \wprod b = \frac{1}{2}\, (ab -
    ba).
\end{equation}
The inner product is the same as the usual ''dot product,'' the only
difference being in the negative sign of the $a_0 b_0$ term; this is
to be expected and is similar to what one finds in special
relativity. The outer product represents an oriented area; in
Euclidean 3-space it can be linked to the "cross product" by the
relation $\mathrm{cross}(\mathbf{a},\mathbf{b}) = - \sigma_{123}
\mathbf{a} \wprod \mathbf{b}$; here we introduced bold characters
for 3-dimensional vectors and avoided defining a symbol for the
cross product because we will not use it again. We also used the
convention that interior and exterior products take precedence over
geometric product in an expression.

When a vector is operated with a multivector the inner product
reduces the grade of each element by one unit and the outer product
increases the grade by one. We will generalize the definition of
inner and outer products below; under this generalized definition
the inner product between a vector and a scalar produces a vector.
Given a multivector $a$ we refer to its grade-$r$ part by writing
$<\!a\!>_r$; the scalar or grade zero part is simply designated as
$<\!a\!>$. By operating a vector with itself we obtain a scalar
equal to the square of the vector's length
\begin{equation}
    a^2 = aa = a \dprod a + a \wprod a = a \dprod a.
\end{equation}
The definitions of inner and outer products can be extended to
general multivectors
\begin{eqnarray}
    a \dprod b &=& \sum_{\alpha,\beta} \left<<\!a\!>_\alpha \;
    <\!b\!>_\beta \right>_{|\alpha-\beta|},\\
    a \wprod b &=& \sum_{\alpha,\beta} \left<<\!a\!>_\alpha \;
    <\!b\!>_\beta \right>_{\alpha+\beta}.
\end{eqnarray}
Two other useful products are the scalar product, denoted as
$<\!ab\!>$ and commutator product, defined by
\begin{equation}
    a \times b = ab - ba.
\end{equation}
In mixed product expressions we will use the convention that inner
and outer products take precedence over geometric products.

We will encounter exponentials with multivector exponents; two
particular cases of exponentiation are specially important. If $u$
is such that $u^2 = -1$ and $\theta$ is a scalar
\begin{eqnarray}
   \mathrm{e}^{u \theta} &=& 1 + u \theta -\frac{\theta^2}{2!} - u
    \frac{\theta^3}{3!} + \frac{\theta^4}{4!} + \ldots  \nonumber \\
    &=& 1 - \frac{\theta^2}{2!} +\frac{\theta^4}{4!}- \ldots \{=
    \cos \theta \} \nonumber \\
    && + u \theta - u \frac{\theta^3}{3!} + \ldots \{= u \sin
    \theta\} \\
    &=&  \cos \theta + u \sin \theta. \nonumber
\end{eqnarray}
Conversely if $h$ is such that $h^2 =1$
\begin{eqnarray}
    \mathrm{e}^{h \theta} &=& 1 + h \theta +\frac{\theta^2}{2!} + h
    \frac{\theta^3}{3!} + \frac{\theta^4}{4!} + \ldots  \nonumber \\
    &=& 1 + \frac{\theta^2}{2!} +\frac{\theta^4}{4!}+ \ldots \{=
    \cosh \theta \} \nonumber \\
    && + h \theta + h \frac{\theta^3}{3!} + \ldots \{= h \sinh
    \theta\}  \\
    &=&  \cosh \theta + h \sinh \theta. \nonumber
\end{eqnarray}
The exponential of bivectors is useful for defining rotations; a
rotation of vector $a$ by angle $\theta$ on the $\sigma_{12}$ plane
is performed by
\begin{equation}
    a' = \mathrm{e}^{\sigma_{21} \theta/2} a
    \mathrm{e}^{\sigma_{12} \theta/2}= \tilde{R} a R;
\end{equation}
the tilde denotes reversion and reverses the order of all products.
As a check we make $a = \sigma_1$
\begin{eqnarray}
    \mathrm{e}^{-\sigma_{12} \theta/2} \sigma_1
    \mathrm{e}^{\sigma_{12} \theta/2} &=&
    \left(\cos \frac{\theta}{2} - \sigma_{12}
    \sin \frac{\theta}{2}\right) \sigma_1 \nonumber \\
    &&\ast \left(\cos \frac{\theta}{2} + \sigma_{12} \sin
    \frac{\theta}{2}\right)  \\
    &=& \cos \theta \sigma_1 + \sin \theta \sigma_2. \nonumber
\end{eqnarray}
Similarly, if we had made $a = \sigma_2,$ the result would have been
$-\sin \theta \sigma_1 + \cos \theta \sigma_2.$

If we use $B$ to represent a bivector whose plane is normal to
$\sigma_0$ and define its norm by $|B| = (B \tilde{B})^{1/2},$ a
general rotation in 4-space is represented by the rotor
\begin{equation}
    R \equiv e^{-B/2} = \cos\left(\frac{|B|}{2}\right) -  \frac{B}{|B|}
    \sin\left(\frac{|B|}{2}\right).
\end{equation}
The rotation angle is $|B|$ and the rotation plane is defined by
$B.$ A rotor is defined as a unitary even multivector (a multivector
with even grade components only) which squares to unity; we are
particularly interested in rotors with bivector components. It is
more general to define a rotation by a plane (bivector) then by an
axis (vector) because the latter only works in 3D while the former
is applicable in any dimension. When the plane of bivector $B$
contains $\sigma_0$, a similar operation does not produce a rotation
but produces a boost instead. Take for instance $B =  \sigma_{01}
\theta/2$ and define the transformation operator $T = \exp( B)$; a
transformation of the basis vector $\sigma_0$ produces
\begin{eqnarray}
    a' &=& \tilde{T} \sigma_0 T = \mathrm{e}^{-\sigma_{01}\theta/2}
    \sigma_0 \mathrm{e}^{\sigma_{01}\theta/2} \nonumber \\
    &=& \left(\cosh \frac{\theta}{2} - \sigma_{01}
    \sinh \frac{\theta}{2}\right) \sigma_0 \nonumber \\
    &&\ast\left(\cosh \frac{\theta}{2} + \sigma_{01} \sinh
    \frac{\theta}{2}\right) \\
    &=& \cosh \theta \sigma_0 + \sinh \theta \sigma_1. \nonumber
\end{eqnarray}

\section{Time derivative of a 4-dimensional vector}
If there is a refractive index the wave displacement vector can be
written as
\begin{equation}
\label{eq:wavedisp}
    \mathrm{d}x = g_\alpha \mathrm{d}x^\alpha = {n^\beta}_\alpha
    \sigma_\beta \mathrm{d}x^\alpha.
\end{equation}
Because this vector is nilpotent, by virtue of Eq.\
(\ref{eq:nullspecial}), the five coordinates are not independent and
we can divide both members by $\mathrm{d}x^0 = \mathrm{d}t$ defining
the nilpotent vector
\begin{equation}
\label{eq:dotx}
    \dot{x} = g_0 +  g_i \dot{x}^i = {n^\alpha}_0   \sigma_\alpha
    + {n^\beta}_i \sigma_\beta \dot{x}^i.
\end{equation}

Suppose we have a 5D vector $a = \sigma_\alpha a^\alpha$ and we want
to find its time derivative along a path parameterized by $t$, that
is all the $x^i$ are functions of $t$. We can write
\begin{equation}
 \dot{a} = \partial_\beta a^\alpha \dot{x}^\beta \sigma_\alpha;
\end{equation}
where naturally $\dot{x}^0 = 1$. Remembering the definition of
covariant derivative (\ref{eq:covariant}) and Eq.\ (\ref{eq:dotx})
we can modify this equation to
\begin{equation}
\label{eq:timeder}
    \dot{a} = \dot{x}^\beta g_\beta \dprod g^\beta \partial_\beta a^\alpha
    \sigma_\alpha
    = \dot{x} \dprod (\mathrm{D}  a).
\end{equation}
We have expressed vector $a$ in terms of the orthonormed frame in
order to avoid vector derivatives but the result must be independent
of the chosen frame.

This procedure has an obvious dual, which we arrive at by defining
\begin{equation}
    \check{x} = g_\mu \check{x}^\mu + g_4.
\end{equation}
The proper time derivative of vector $a$ is then
\begin{equation}
    \check{a} = \check{x} \dprod (\mathrm{D} a).
\end{equation}

\end{appendix}

% Create the reference section using BibTeX:
\bibliography{Abrev,aberrations,assistentes}

\end{document}